\documentclass[12pt,a4paper]{article}

\usepackage{graphicx}
\usepackage{cite}

\begin{document}

\title{
\vspace*{-1cm}
{\small\sf \rightline{HD-THEP-00-51}}
\vspace*{1.5cm}
{\LARGE\bf Charm quark mass from QCD sum rules for the charmonium system \\[5mm]}
}

\author{
{\Large  M. Eidem\"uller and M. Jamin} \\[5mm]
{\normalsize\sl Institut f\"ur Theoretische Physik, Universit\"at Heidelberg,} \\
{\normalsize\sl Philosophenweg 16, 69120 Heidelberg, Germany}\\
}

\date{}

\maketitle
\thispagestyle{empty}

\vspace{5mm}

\normalsize      

\begin{abstract}
\noindent
In this work, the charm quark mass is obtained
from a QCD sum rule analysis of the charmonium system. 
In our investigation we include results from 
nonrelativistic QCD at next-to-next-to-leading order.
Using the pole mass scheme, we obtain a value of
$M_c=1.70\pm 0.13$ GeV for the charm pole mass. The introduction
of a potential-subtracted mass leads to an improved scale
dependence. The running ${\rm \overline{MS}}$-mass is then determined to be
$m_c(m_c) = 1.23 \pm 0.09$ GeV.

\vspace{20mm}

\end{abstract}

\newpage
\setcounter{page}{1}

\newcommand{\nn}{\nonumber}
\newcommand{\mev}{\mbox{\rm MeV}}
\newcommand{\gev}{\mbox{\rm GeV}}
\newcommand{\eqn}[1]{(\ref{#1})}
\newcommand{\MSb}{{\overline{MS}}}
\newcommand{\ep}{\epsilon}
\newcommand{\IM}{\mbox{\rm Im}}

\newcommand{\jhep}[3]{{\it JHEP }{\bf #1} (#2) #3}
\newcommand{\nc}[3]{{\it Nuovo Cim. }{\bf #1} (#2) #3}
\newcommand{\npb}[3]{{\it Nucl. Phys. }{\bf B #1} (#2) #3}
\newcommand{\npps}[3]{{\it Nucl. Phys. }{\bf #1} {\it(Proc. Suppl.)} (#2) #3}
\newcommand{\plb}[3]{{\it Phys. Lett. }{\bf B #1} (#2) #3}
\newcommand{\pr}[3]{{\it Phys. Rev. }{\bf #1} (#2) #3}
\newcommand{\prd}[3]{{\it Phys. Rev. }{\bf D #1} (#2) #3}
\newcommand{\prl}[3]{{\it Phys. Rev. Lett. }{\bf #1} (#2) #3}
\newcommand{\prep}[3]{{\it Phys. Rep. }{\bf #1} (#2) #3}
\newcommand{\rpp}[3]{{\it Rept. Prog. Phys. }{\bf #1} (#2) #3}
\newcommand{\zpc}[3]{{\it Z. Physik }{\bf C #1} (#2) #3}
\newcommand{\sjnp}[3]{{\it Sov. J. Nucl. Phys. }{\bf #1} (#2) #3}
\newcommand{\jetp}[3]{{\it Sov. Phys. JETP }{\bf #1} (#2) #3}
\newcommand{\jetpl}[3]{{\it JETP Lett. }{\bf #1} (#2) #3}
\newcommand{\ijmpa}[3]{{\it Int. J. Mod. Phys. }{\bf A #1} (#2) #3}
\newcommand{\hepph}[1]{{\tt hep-ph/#1}} 
\newcommand{\hepth}[1]{{\tt hep-th/#1}} 
\newcommand{\heplat}[1]{{\tt hep-lat/#1}}


\section{Introduction}

Quantum Chromo Dynamics (QCD), describing the strong interactions, 
is one of the key components of the Standard Model. 
The determination of its parameters remains an essential
task within modern particle physics.
Much effort has therefore been put into the study of quark masses.
However, in most cases where the system is sensitive to 
mass effects, also confinement plays an important role.
Apart from the top mass, the quark masses can thus only be 
calculated by nonperturbative methods like
QCD sum rules \cite{svz:79,rry:85,n:89}, 
lattice QCD \cite{r:92,mm:94}
or chiral perturbation theory \cite{gl:85,p:95}.

In the past, QCD moment sum rule
analyses have been successfully applied for extracting the charm and
bottom quark masses from experimental data on the charmonium and 
bottonium systems respectively.
The fundamental quantity in these investigations is the 
vacuum polarisation function $\Pi(q^2)$:
\begin{equation}
  \label{eq:a}
  \Pi_{\mu\nu}(q^2) = i \int d^4 x \ e^{iqx}\, \langle T\{j_\mu(x) j_\nu(0)\}\rangle
  = (q_\mu q_\nu-g_{\mu\nu}q^2)\,\Pi(q^2)\,,
\end{equation}
where in the charm case the vector current is represented by 
$j_\mu(x)=(\bar{c}\gamma_\mu c)(x)$.
Via the optical theorem, the experimental cross 
section $\sigma(e^+ e^- \to c\bar{c})$
is related to the imaginary part of $\Pi(s)$:
\begin{equation}
  \label{eq:b}
  R(s)=\frac{1}{Q_c^2}\,\frac{\sigma(e^+ e^- \to c\bar{c})}
  {\sigma(e^+ e^- \to \mu^+ \mu^-)}=12\pi\,\IM\, \Pi(s+i\ep)\,.
\end{equation}
Using a dispersion relation, we can express the moments ${\cal M}_{n}$ 
of $\Pi(s)$ by integrals over the spectral density $R(s)$:
\begin{equation}
  \label{eq:c}
 {\cal M}_{n} = \frac{12\pi^2}{n!} \left(4m^2 \frac{d}{ds}\right)^n
  \Pi(s)\bigg|_{s=0} 
 =\left(4m^2\right)^n 
 \int_{s_{min}}^\infty \!ds\,\frac{R(s)}{s^{n+1}}\,.
\end{equation}
For convenience, we have defined the moments as dimensionless quantities.
In addition, it will prove useful to express the moments by 
integrals over the velocity $v=\sqrt{1-4m^2/s}$: 
\begin{equation}
  \label{eq:d}
 {\cal M}_{n} = 2 \int_0^1 \!dv \, v(1-v^2)^{n-1}R(v)\,.
\end{equation}

The moments can either be calculated theoretically, including 
perturbation theory, Coulomb resummation and nonperturbative contributions,
or can be obtained from experiment. In this way, we can relate the charm
quark mass to the hadronic properties of the charmonium system.

During the last years, several analyses for the bottonium system
have been performed \cite{jp:97,v:95,kpp:98,my:98,pp:99,b:99:2,h:99,h:00}.
In these investigations, it turned out that the largest theoretical
contributions arise from the threshold expansion in the framework
of nonrelativistic QCD (NRQCD) which is known up to
next-to-next-to-leading order (NNLO) \cite{pp:99,h:99,py:98}. 
The correlator can be expressed
in terms of a Greens function, which shows a continuous spectrum
above threshold, but also contains poles below threshold.

Yet these new theoretical developments have not
been applied to the charmonium system. Due to the low scales involved, 
the analysis becomes more delicate here. We have therefore introduced
a new method of analysis which allows for investigation of the influence
of the individual contributions on the error. In particular, 
we have taken special care in obtaining the pole contributions 
and the reconstruction of the spectral density above threshold.

So far, we have not specified the mass definition to be used in 
eq. \eqn{eq:c}.
A natural choice for this mass is the pole mass $M$.
In the first part of our numerical analysis, we will use the pole mass
scheme to extract the charm pole mass. However, as the pole mass suffers
from renormalon ambiguities \cite{b:99}, it can only be
determined up to corrections of order $\Lambda_{QCD}$. 
In the second part of
our analysis we shall therefore use the recently introduced 
potential-subtracted mass $m_{PS}$ \cite{b:98}.
From this mass definition we can obtain the ${\rm \MSb}$-mass
more accurately than from the pole mass scheme.

In the next section, we shall present the contributions from
the threshold expansion in the framework of NRQCD. The
perturbative expansion which is needed for the reconstruction
of the spectral density will be discussed in section 3.
Afterwards, we will shortly describe the nonperturbative
contributions and the phenomenological spectral function.
In the numerical analysis, we shall first explain the method
of analysis.
We will then obtain the pole mass
and the ${\rm \MSb}$-mass from an analysis
of the pole mass and the PS-mass scheme respectively. 
The origin of different contributions to the error will be carefully
investigated. We shall conclude with a summary and an 
outlook.

 
\section{Coulomb resummation}

Close to threshold, it turns out that the relevant expansion
parameter is $\alpha_s/v$ rather than $\alpha_s$ 
since the velocity
becomes of order $\alpha_s$. These terms can be resummed 
in the framework of NRQCD and give the largest contributions 
to the theoretical moments.
 
The correlator is then expressed in terms of a Greens function 
\cite{pp:99,ps:91}:
\begin{equation}
  \label{eq:e}
  \Pi(s)=\frac{N_c}{2M^2}\left(C_h(\alpha_s)G(k)+\frac{4k^2}{3M^2}G_C(k)\right)\,,
\end{equation}
where $k=\sqrt{M^2-s/4}$ and $M$ represents the pole mass.
The constant $C_h(\alpha_s)$ is a perturbative coefficient which is
needed for the matching between the full and the nonrelativistic 
theory. It naturally depends on the hard scale.
The Greens function is analytically known up to NNLO \cite{pp:99}
and sums up terms of order
$\alpha_s^n/v^{n-k}$ for $n\geq 0$ and $k=1,2,3$.
It is crucial for the analysis that the result depends on three scales.
While the hard scale $\mu_{hard}$ is responsible
for the hard perturbative processes,
the soft scale $\mu_{soft}$ governs the expansion of the
Greens function. Furthermore, the factorisation scale
$\mu_{fac}$ separates the contributions of large and small 
momenta and plays the role of an infrared cutoff.

The Greens function contains two parts: 
the continuum above and the poles below threshold respectively.
We are interested in both contributions separately: first, the individual
corrections can be analysed and their error estimated. Second, in our
numerical analysis we will reconstruct the spectral density above 
threshold and we thus need the corresponding spectral density 
at low velocities.
In principle, the expressions for the energies and decay widths of the
poles have been calculated.
However, in the actual case of the charm quark the
expansion does not converge well. We will therefore choose a different
method of evaluation \cite{e:00,ej:00}. 
Since the expansion for the Greens function 
at NNLO is known analytically \cite{pp:99}, 
we can evaluate their contribution to the moments
numerically by performing the derivatives.
On the other hand, by using a dispersion relation, we can derive
the continuum from the imaginary part of the correlator. 
From the difference we can then obtain the pole contributions:
\begin{equation}
  \label{eq:f}
  {\cal M}_n^{Poles} = \frac{12 \pi^2}{n!}\left(4M^2\frac{d}{ds}\right)^n
    \Pi(s)\Big|_{s=0}
    -12 \pi \left(4M^2\right)^n \int_{4M^2}^\infty ds\  
    \frac{\IM \,\Pi(s)}{s^{n+1}}\,.
\end{equation}
Since we will not evaluate the poles near threshold, but rather calculate
their contributions to the moments in a region where perturbation theory
is expected to be valid, the convergence of the pole contributions
is improved. Nevertheless, the poles will give the largest contribution
to the theoretical moments and thus the dependence on the
scales will remain relatively strong. 
In the numerical analysis we will give a detailed account on
the size and behaviour of these contributions.

The large corrections are partly due to the definition of the pole mass. 
These contributions can
be reduced by using an intermediate mass definition. In this analysis
we will use the so-called potential-subtracted (PS) mass \cite{b:98} where
the potential below a separation scale $\mu_{sep}$ is subtracted:
\begin{equation}
  \label{eq:g} 
  m_{PS}(\mu_{sep}) = M-\delta  m(\mu_{sep})\,,\quad
  \delta m(\mu_{sep}) = -\frac{1}{2}\int\limits_{|{\bf q}|<\mu_{sep}}
  \!\!\!\frac{d^3 q}{(2\pi)^3}\,V(q)\,.
\end{equation}
Since the QCD potential contains the same renormalon ambiguities
as the pole mass, in the difference they are cancelled.
This mass definition thus 
leads to an improved scale dependence and a more precise
determination of the ${\rm \MSb}$-mass.


\section{Perturbative expansion}

The perturbative spectral function $R(s)$ can be expanded in powers of the
strong coupling constant $\alpha_s$,
\begin{equation}
  \label{eq:h}
  R^{Pt}(s)= R^{(0)}(s)+ \frac{\alpha_s}{\pi}\,R^{(1)}(s)+ 
  \frac{\alpha_s^2}{\pi^2} \,R^{(2)}(s)+\ldots\,.
\end{equation}
From this expression the corresponding moments
${\cal M}_n^{Pt}$ can be calculated via the integral of eq. \eqn{eq:d}.
The first two terms are known analytically and can for example be
found in ref. \cite{jp:97}. $\Pi^{(2)}(s)$ is still not fully known
analytically. However, the method of Pad{\'e}-approximants has been
exploited to calculate $\Pi^{(2)}$ numerically, using available
results at high energies, analytical results for the first eight
moments and the known threshold behaviour 
\cite{cks:96,cks:97}.
This information is sufficient to obtain a numerical
approximation of $\Pi^{(2)}(s)$ in the full energy range.

The numerical stability of the results can be checked in different 
ways. By choosing different Pad{\'e}-approximants or by selecting
a smaller set of input data the results for the moments remain
almost unchanged. Furthermore, some contributions to the spectral
density like those from internal quark loops are known analytically
and are in very good agreement with the numerical spectral density.
This provides strong support 
that the numerically obtained spectral density
gives a very good approximation to the exact spectral density.


\section{Condensate contributions}

The nonperturbative effects on the vacuum correlator are parametrised
by the condensates. The leading correction is the gluon condensate
contribution which is known up to next-to-leading order \cite{bbifts:94}.
Furthermore, the dimension 6 and 8 contributions have been calculated
and will be included in our analysis \cite{nr:83:1,nr:83:2,bg:85}.
In the analysis below, we have employed a value of 
$\langle \alpha_s/\pi FF\rangle = 0.024\pm 0.012 \ \gev^4$ 
for the gluon condensate.

It will turn out that the condensate contributions are suppressed
when compared to former charmonium sum rule analyses \cite{dgp:94,n:94:2}
and only have little influence on the mass. 
Besides an increase of the theoretical moments from the
Coulomb contributions we
will restrict the moments to $n\leq 7$ where the nonperturbative contributions
are relatively small. Since we obtain a larger pole mass
than the former analyses, the condensates,
starting with a power of $1/M^4$, are suppressed further.


\section{Phenomenological spectral function}

Experimentally, the lowest lying six $\psi$-resonances have been observed.
Since the widths of the poles are very small compared to the masses,
the narrow-width approximation provides an excellent description
of these states. To model the contributions of the hadronic continuum,
we use the assumption of quark-hadron-duality and integrate over the
perturbative spectral density:
\begin{equation}
  \label{eq:i}
  \frac{{\cal M}_{n}}{(4M^2)^n} = \frac{9\pi}{\alpha_{em}^2 Q_c^2}
  \sum_{k=1}^6 \frac{\Gamma_k}{E^{2n+1}_k}
  + \int_{s_0}^\infty ds\,\frac{R^{Pt}(s)}{s^{n+1}}\,.
\end{equation}
To estimate the continuum contribution we will use a threshold
in the range of $3.6$ GeV $\leq \sqrt{s_0}\leq 4.2$ GeV with a central
value of $\sqrt{s_0}= 3.8$ GeV. 
The lower value corresponds to using a perturbative continuum
around the $\psi(2S)$ but nevertheless including
the higher resonance contributions. This is certainly over-counting
the continuum contribution. The higher value would assume
no additional continuum contribution until the $\psi(5S)$
resonance.
However, the most dominant phenomenological
contributions come from the first two $\psi$-resonances resulting
in a small influence of the continuum even for low values of $n$.


\section{Numerical analysis}


\subsection{Method of analysis}

Besides the contributions from the poles of the Greens function 
and the condensates, the theoretical part of the correlator
contains the spectral density above threshold. Now we will discuss
the different parts of the spectral density.
A more detailed description of our procedure will be presented
in a forthcoming publication \cite{ej:inprep}.

For high velocities the spectral density is well described
by the perturbative expansion where we have fixed the hard
scale to the pole mass. However, as one approaches smaller
values of $v$, the perturbative expansion breaks down. The resummed
spectral density, on the other hand, gives a good description
for low values of $v$, but becomes unreliable for high velocities.
For these reasons, we will introduce a separation velocity $v_{sep}\approx 0.3$.
Above $v_{sep}$ we will use the perturbative spectral density. Below
$v_{sep}$, we take the resummed spectral density adding the 
terms which are included in perturbation theory but not in resummation.

\begin{figure}
\begin{center}
\includegraphics[width=11cm,height=7cm]{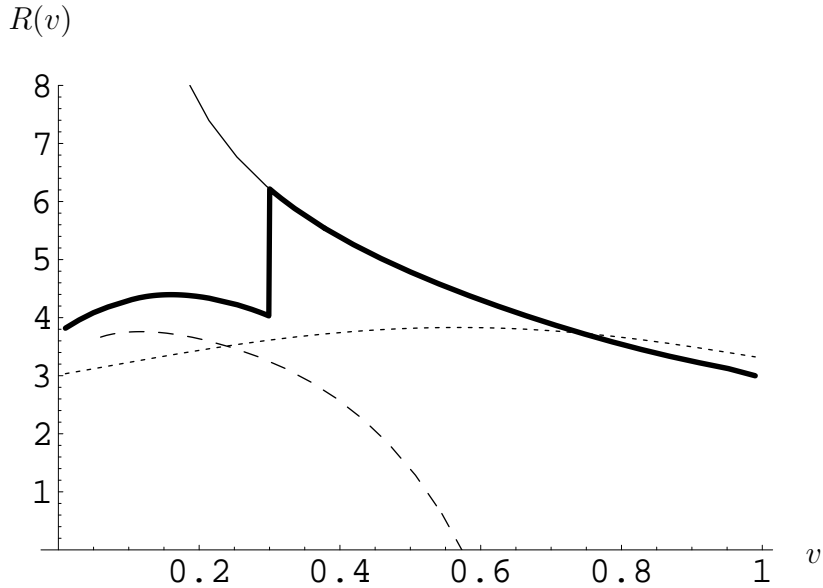}
\put(-10.7,7.5){$R(v)$}
\put(-.1,.4){$v$}
\caption{\label{fig:a}Thick solid line: reconstructed spectral density;
Thin solid line: perturbative spectral density;
dotted  line: perturbation theory at NLO;
dashed line: resummed spectral density.}
\end{center}
\end{figure}
In fig. \ref{fig:a} we have displayed the different contributions. The dotted
line represents the perturbative expansion at NLO. The thin
solid line also includes the NNLO. Whereas for high velocities
the perturbative expansion is well convergent, the importance
of the higher corrections increases for smaller $v$. The dashed line
is the resummed spectral density and the thick solid line the 
reconstructed spectral density. For the charmonium system, there
exists a range of intermediate values of $v$ where neither the
perturbative expansion nor the resummation can be trusted. 
Indeed, it can be clearly seen that the reconstructed
spectral density shows a gap at the separation velocity.
To estimate the error we have varied $v_{sep}$ between 0.2 and 0.4.
The analysis shows that though the introduction of the separation
velocity stabilises the sum rules, the variation only has a minor 
influence on the mass.


\subsection{Pole mass scheme}

Since the threshold region becomes more important for larger 
values of $n$, as can already be seen from the factor $(1-v^2)^{n-1}$
of eq. \eqn{eq:d}, also the NLO and NNLO corrections grow
for larger $n$. We will therefore restrict our analysis to 
moments with $n\leq 7$. As our method of analysis
needs moments of $n\geq 3$ to reconstruct the spectral density,
we will use $3\leq n\leq 7$. 
Since the continuum part of the phenomenological spectral density is relatively
small even for low values of $n$, it does not set a lower bound on $n$.
As central values for our scales we have selected: 
\begin{equation}
  \label{eq:j}
  \mu_{soft}= 1.2 \ \mbox{GeV}\,, \quad \mu_{fac}= 1.45\ \mbox{GeV}\,,\quad
  \mu_{hard}= 1.7\ \mbox{GeV}\,. 
\end{equation}
We have set the hard scale equal to the central value for the pole mass.
For the soft scale, we would have liked to choose a 
value around $\mu_{soft}\sim 0.8-0.9$ GeV, 
but the NNLO corrections get out of control for these scales.
Thus, we take a central value of $\mu_{soft}= 1.2$ GeV
and vary it in an interval of $1.05\ \gev \leq \mu_{soft}\leq 1.5\ \gev$.
The factorisation scale separates the different regions and should
lie between the two other scales.
\begin{figure}
\begin{center}
\includegraphics[width=11cm,height=7cm]{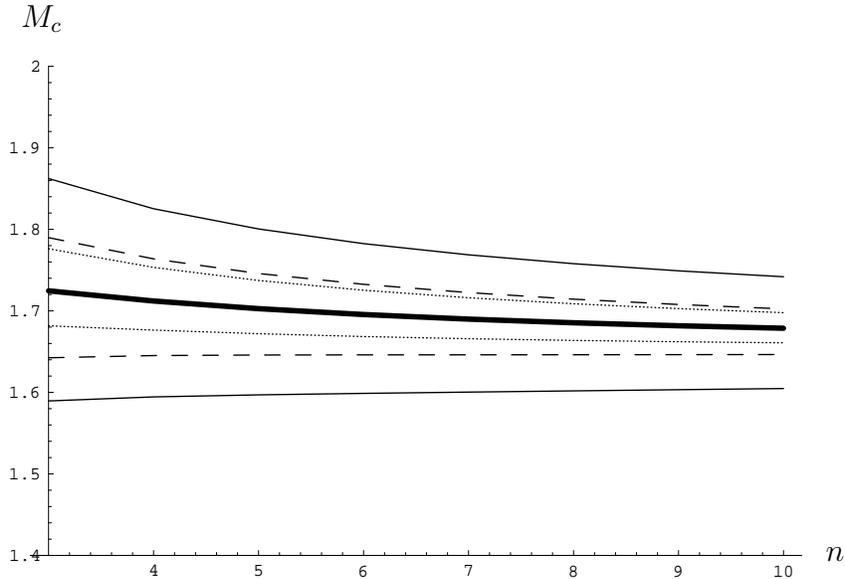}
\put(-10.6,7.4){$M_c$}
\put(0.1,.3){$n$}
\caption{\label{fig:b}
Thick solid line: central pole mass;
thin solid lines: $M_c$ for  $\mu_{soft}=1.05$
and 1.5 GeV;
dashed lines: $M_c$ for  $\mu_{fac}=1.2$
and 1.7 GeV;
dotted lines: $M_c$ for  $\mu_{hard}=1.4$
and 2.5 GeV.}
\end{center}
\end{figure}

\begin{table}
\begin{center}
\begin{tabular}{c|c|c|c|c|c|}
$n$  & 3 & 4 & 5 & 6 & 7 \\ \hline
${\cal M}_n^{Poles}$ & 3.84 & 4.44 & 5.11 & 5.87 & 6.73  \\ 
${\cal M}_n^{Pt}$ & 1.92 & 1.55 & 1.32 & 1.15 & 1.03  \\ 
${\cal M}_n^{NRResum}$ & 0.28 & 0.27 & 0.26 & 0.24 & 0.23  \\
${\cal M}_n^{Counter}$ & 0.69 & 0.66 & 0.64 & 0.62 & 0.60  \\ 
${\cal M}_n^{Continuum}$ & 0.75 & 0.47 & 0.31 & 0.21 & 0.15  \\
${\cal M}_n^{Condensates}$ & -0.03 & -0.05 & -0.06 & -0.08 & -0.09  \\ \hline
\end{tabular}
\caption{\label{tab:a}Moments for different $n$ with the parameters
$\mu_{soft} = 1.2$ GeV, $\mu_{fac}= 1.45$ GeV,
$\mu_{hard}= 1.7$ GeV,
$\sqrt{s_0}= 3.8$ GeV and $v_{sep}=0.3$.}
\end{center}
\end{table}

In table \ref{tab:a} we have collected the individual 
moments for different values of $n$. 
${\cal M}_n^{Poles}$ are the theoretical poles of the Greens function,
${\cal M}_n^{NRResum}$ the moments from the nonrelativistic spectral
density and ${\cal M}_n^{Counter}$ contains the double counted terms
from the perturbative expansion below $v_{sep}$.
In this scheme the theoretical moments are dominated by the pole 
contributions. Though the size of these terms grow with larger $n$,
the mass remains relatively stable since the contributions
of the moments to the mass is suppressed by a power of $1/2n$.

The thick solid line in fig. \ref{fig:b} shows the central value
for the pole mass $M_c=1.70$ GeV with the parameters from eq. \eqn{eq:j}.
The error is dominated by the variation of the scales. We obtain:
\begin{eqnarray}
  \label{eq:k}
  1.05 \ \mbox{GeV} \leq \mu_{soft}\leq 1.5\ \mbox{GeV}:&&\qquad
  \Delta M_c=100\ \mbox{MeV}\,, \nn\\
  1.2 \ \mbox{GeV} \leq \mu_{fac\ }\leq 1.7\ \mbox{GeV}:&&\qquad
  \Delta M_c=50\ \mbox{MeV}\,, \nn\\
  1.4 \ \mbox{GeV} \leq \mu_{hard}\leq 2.5\ \mbox{GeV}:&&\qquad
  \Delta M_c=40\ \mbox{MeV}\,. 
\end{eqnarray}
\begin{table}
\begin{center}
\begin{tabular}{c|c|c|c|c|c|c|}\hline
$\mu_{soft}$  & 1.05 & 1.1 & 1.2 & 1.3 & 1.4 & 1.5  \\ 
${\cal M}_5^{Poles}$ & 9.50 & 7.52 & 5.14 & 3.80 & 2.97 & 2.41 \\ \hline
$\mu_{fac}$  & 1.2 & 1.3 & 1.4 & 1.5 & 1.6 & 1.7  \\
${\cal M}_5^{Poles}$ & 6.79 & 6.12 & 5.46 & 4.82 & 4.18 & 3.55  \\ \hline
$\mu_{hard}$  & 1.4 & 1.5 & 1.6 & 1.8 & 2.0 & 2.5  \\ 
${\cal M}_5^{Poles}$ & 4.24 & 4.62 & 4.93 & 5.45 & 5.85 & 6.56 \\ \hline
\end{tabular}
\caption{\label{tab:b}Pole moments with $n=5$ for different
$\mu_{soft}$ with
$\mu_{fac}=1.45$ GeV and $\mu_{hard}=1.7$ GeV,
for different $\mu_{fac}$ with $\mu_{soft}=1.2$ GeV and
$\mu_{hard}=1.7$ GeV and
for different $\mu_{hard}$ with $\mu_{soft}=1.2$ GeV and
$\mu_{fac}=1.45$ GeV.} 
\end{center}
\end{table}
In particular, in table \ref{tab:b} we have listed the influence
of the scales on the pole contributions. We have varied the scales
within physically reasonable ranges. Since the convergence
of the nonrelativistic expansion is not very good for the 
charmonium system, the scales cannot be chosen arbitrarily 
far away from their central values.
Though the analysis is stable inside the given intervals,
the expressions tend to become unstable for scales outside of the
chosen ranges.
\begin{table}
\begin{center}
\begin{tabular}{c||c|c|c|c|c|c|c|c|}
$\mu_{soft}$ & & 1.05 & 1.1 & 1.2 & 1.3 & 1.4 & 1.5  \\ \hline 
 & LO & 3.03 & 2.57 & 1.95 & 1.56 & 1.29 & 1.10  \\ 
${\cal M}_5^{Poles}$ & NLO & 5.60 & 4.75 & 3.61 & 2.89 & 2.40 & 2.04  \\
 & NNLO & 9.50 & 7.52 & 5.14 & 3.80 & 2.97 & 2.41  \\ \hline
 & LO & 0.64 & 0.61 & 0.57 & 0.54 & 0.51 & 0.49  \\ 
${\cal M}_5^{NRResum}$ & NLO & 0.24 & 0.27 & 0.29 & 0.30 & 0.31 & 0.31  \\
 & NNLO & 0.31 & 0.28 & 0.23 & 0.20 & 0.18 & 0.17  \\ \hline
\end{tabular}
\caption{\label{tab:c}Size of the moments
from the  poles and the resummed spectral density 
at LO, NLO and NNLO for different values of $\mu_{soft}$.}
\end{center}
\end{table}
In table \ref{tab:c} we have confronted the NLO and NNLO corrections
to the LO. The NNLO corrections turn out not to be so
large for values of $\mu_{soft} \geq 1.2$ GeV. 
But this may happen accidentally since there are cancellations for 
the combination of scales used in this analysis. Therefore 
this should not be taken as an
indication for small higher order corrections. 
In order to obtain a more conservative estimate,
thus we have used the variation of the scales in our
error analysis.

A significant uncertainty also comes from $\Lambda_{QCD}$.
By choosing $\Lambda_{QCD}=330\pm 30$ MeV we get an error of
$\Delta M_c=50$ MeV.
The variation of the continuum threshold $s_0$ and the error from
the experimentally measured decay widths only have a small influence on the mass.
We have summarised the results in table \ref{tab:d}.
\begin{table}
\begin{center}
\begin{tabular}{|l|r|}\hline
\multicolumn{1}{|c}{Source} & \multicolumn{1}{|c|}{$\Delta M_c$} \\ \hline 
Variation of $\mu_{soft}$ & 100 MeV \\
Variation of $\mu_{fac}$ & 50 MeV \\
Variation of $\mu_{hard}$ & 40 MeV \\ 
Threshold $s_0$ & 10 MeV \\
Experimental error & 15 MeV \\
Variation of $v_{sep}$ & 10 MeV \\
Variation of $\Lambda_{QCD}$ & 50 MeV \\ \hline
Total error & 130 MeV \\ \hline 
\end{tabular}
\caption{\label{tab:d}Single contributions to the error of $M_c$.}
\end{center}
\end{table}
Adding the errors in quadrature we obtain the charm pole mass:
\begin{equation}
  \label{eq:l}
  M_c=1.70 \pm 0.13 \ \mbox{GeV}\,.
\end{equation}
Using the  three-loop relation between the pole and the ${\rm \MSb}$-mass
which has been calculated recently \cite{mr:99,cs:99},
we obtain $m_{c}(m_{c})=1.20\pm 0.13$ GeV for the ${\rm \MSb}$-mass.
In addition, we get an error from the perturbative relation between
the two masses which does not converge for the charm case and
can be of order $\Lambda_{QCD}$.
By using the PS-mass, in the next section we will derive a more
precise value for the ${\rm \MSb}$-mass.


\subsection{Potential-subtracted mass scheme}

First, we have to choose a value for the separation scale $\mu_{sep}$.
This scale should be taken large enough in order to guarantee a perturbative
relation to the ${\rm \MSb}$-mass. On the other hand, it should be 
smaller than $Mv$. Both conditions cannot be well fulfilled
at the same time. As a compromise value we will choose
$\mu_{sep}=1.0\pm0.2$ GeV. In this scheme, the pole contributions
from the Greens function turn out to be smaller than in the 
pole mass scheme. The contributions from the condensates get more
important here and we shall restrict our analysis to $n\leq 6$ to where
these corrections are under good control. 

\begin{figure}
\begin{center}
\includegraphics[width=11cm,height=7cm]{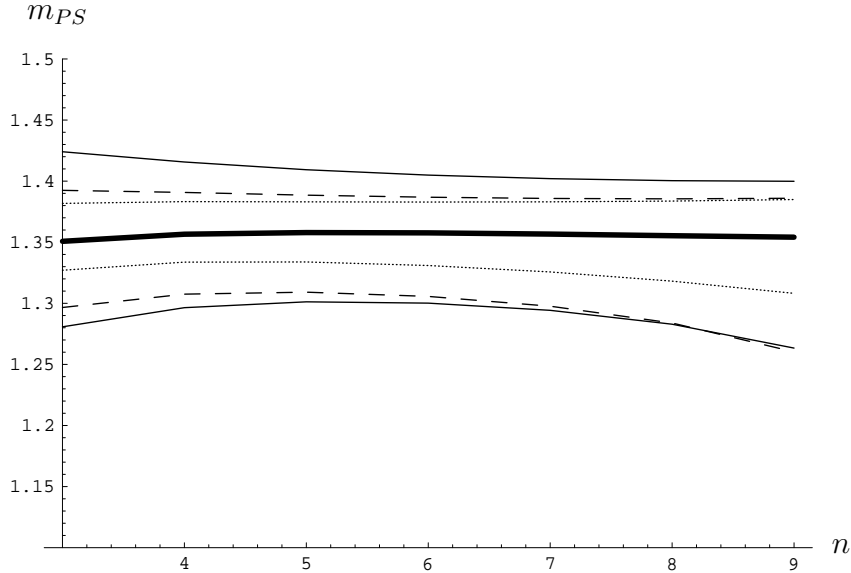}
\put(-10.6,7.4){$m_{PS}$}
\put(0.1,.3){$n$}
\caption{\label{fig:c}
Thick solid line: central PS-mass;
thin solid lines: $m_{PS}$ for  $\mu_{soft}=1.05$
and 1.5 GeV;
dashed lines: $m_{PS}$ for  $\mu_{fac}=1.2$
and 1.7 GeV;
dotted lines: $m_{PS}$ for  $\mu_{hard}=1.4$
and 2.5 GeV.}
\end{center}
\end{figure}
Using the same central 
values for the scales \eqn{eq:j} we obtain 
$m_{PS}(\mu_{sep}=1.0) = 1.35$ GeV and from this value a ${\rm \MSb}$-mass
of $m_{c}(m_{c}) = 1.23$ GeV. The introduction of the intermediate
mass definition leads to a reduced scale dependence:
\begin{eqnarray}
  \label{eq:m}
  1.05 \ \mbox{GeV} \leq \mu_{soft}\leq 1.5\ \mbox{GeV}:&&\qquad 
  \Delta m_{PS}=60\ \mbox{MeV}\,,\nn\\
  1.2 \ \mbox{GeV} \leq \mu_{fac\ }\leq 1.7\ \mbox{GeV}:&&\qquad
  \Delta m_{PS}=40\ \mbox{MeV}\,,\nn\\
  1.4 \ \mbox{GeV} \leq \mu_{hard}\leq 2.5\ \mbox{GeV}:&&\qquad
  \Delta m_{PS}=30\ \mbox{MeV}\,. 
\end{eqnarray}
\begin{table}
\begin{center}
\begin{tabular}{c|c|c|c|c|}
$n$  & 3 & 4 & 5 & 6  \\ \hline
${\cal M}_n^{Poles}$ & 0.59 & 0.40 & 0.27 & 0.19  \\ 
${\cal M}_n^{Pt}$ & 1.16 & 0.83 & 0.63 & 0.49   \\ 
${\cal M}_n^{NRResum}$ & 0.33 & 0.32 & 0.30 & 0.29  \\
${\cal M}_n^{Counter}$ & 0.53 & 0.47 & 0.42 & 0.37   \\
${\cal M}_n^{Continuum}$ & 0.42 & 0.20 & 0.09 & 0.03   \\
${\cal M}_n^{Condensates}$ & -0.07 & -0.09 & -0.12 & -0.14 \\ \hline
\end{tabular}
\caption{\label{tab:e}Moments for different $n$ with the parameters 
$\mu_{soft} = 1.2$ GeV, $\mu_{fac}= 1.45$ GeV, $\mu_{hard}=1.7$ GeV,
$\sqrt{s_0}= 3.8$ GeV and $v_{sep}=0.3$.}
\end{center}
\end{table}
In table \ref{tab:e} we again have collected the moments for
different values of $n$. The influence of the pole contributions is clearly
reduced and becomes even smaller for higher values of $n$. 
Since we have chosen
the separation scale relatively large in comparison to $Mv$,
the PS-mass lies a good deal lower than the pole mass and we are further away
from the threshold region. The convergence of the perturbative
expansion is improved in this scheme as well. 
\begin{table}
\begin{center}
\begin{tabular}{c||c|c|c|c|c|}
$\mu_{sep}$ & 0.8 & 0.9 & 1.0 & 1.1 & 1.2  \\ \hline 
$m_{PS}$ & 1.38 & 1.36 & 1.35 & 1.34 & 1.33 \\ 
$m_{c}(m_c)$ & 1.20 & 1.22 & 1.23 & 1.24 & 1.26 \\ \hline
\end{tabular}
\caption{\label{tab:f}Change of the masses for different
values of  $\mu_{sep}$.}
\end{center}
\end{table}
In table \ref{tab:f} we have varied the separation scale
$\mu_{sep}$ between $0.8 \ \gev\leq\mu_{sep}\leq 1.2\ \gev$.
The error on the ${\rm \MSb}$-mass is about 30 MeV. The change of 
the two masses is not directly correlated
since the variation of $\mu_{sep}$ also changes the relation
between the masses.
When compared to the pole mass scheme, the uncertainty from 
$\alpha_s$ is reduced and
the influence of the condensates
becomes  more important here.
\begin{table}
\begin{center}
\begin{tabular}{|l|r|r|} \hline
\multicolumn{1}{|c}{Source} & \multicolumn{1}{|c|}{$\Delta m_{PS}$} 
 & \multicolumn{1}{c|}{$\Delta m_{c}$} 
\\ \hline 
Variation of $\mu_{soft}$ & 60 MeV & 50 MeV \\
Variation of $\mu_{fac}$ & 40 MeV  & 40 MeV\\
Variation of $\mu_{hard}$ & 30 MeV& 30 MeV \\ 
Variation of $\mu_{sep}$ & 30 MeV  & 30 MeV\\
Threshold $s_0$ & 30 MeV & 30 MeV\\
Experimental error & 10 MeV & 10 MeV\\
Condensates & 20 MeV & 20 MeV\\
Variation of $v_{sep}$ & 10 MeV & 10 MeV\\
Variation of $\Lambda_{QCD}$ & 10 MeV& 20 MeV \\ \hline
Total error & 90 MeV& 90 MeV \\ \hline 
\end{tabular}
\caption{\label{tab:g}Single contributions to the error of $m_{PS}$ and $m_{c}$.}
\end{center}
\end{table}

In table \ref{tab:g} we have listed the individual contributions
to the error of $m_{PS}$ and $m_{c}$. Finally, we
obtain the masses:
\begin{eqnarray}
  \label{eq:n}
  m_{PS}(\mu_{sep}=1.0) &=& 1.35 \pm 0.09\ \mbox{GeV}\,,\nn\\
  m_{c}(m_{c}) &=& 1.23 \pm 0.09 \ \mbox{GeV}\,.
\end{eqnarray}
When we compare this value of the ${\rm \MSb}$-mass to the one
from the last section, $m_{c}(m_{c})=1.20$ GeV,
we see that the central value remains almost unchanged. 
This is not self-evident, as the dominating pole contributions 
are reduced in the PS-scheme and the relative influence 
of the individual contributions is shifted.


\section{Conclusions}

To summarise, our values obtained for the charm quark
pole- and ${\rm \MSb}$-mass are as follows:
\begin{eqnarray}
  \label{eq:n:2}
  M_c &=& 1.70 \pm 0.13 \ \mbox{GeV}\,,\nn\\ 
  m_{c}(m_{c}) &=& 1.23 \pm 0.09 \ \mbox{GeV}\,.
\end{eqnarray}
The obtained value for the pole mass lies somewhat higher
than in previous sum rule analyses \cite{dgp:94,n:94:2}.
In  \cite{dgp:94} the authors used perturbation theory to NLO
resulting in a value of $M_c=1.46\pm 0.07$ GeV. In the second investigation
\cite{n:94:2,n:98} the analysis has been performed
in the ${\rm \MSb}$-scheme with perturbation theory to NLO. 
Using the NLO relation to the pole mass the author obtains
$m_{c}(m_{c}) = 1.26\pm 0.05$ GeV and $M_c = 1.42 \pm 0.03$ GeV.
The author has also performed an analysis using resummation
in LO with a value of $M_c = 1.45 \pm 0.07$ GeV.

In our analysis the increased value of the pole mass is essentially
due to large Coulomb contributions which have not been included in
former analyses. As a consequence, the error becomes larger as well. 
In a recent analysis, the pole mass has been estimated from the charmonium
ground state at NNLO \cite{py:98}. Here the authors obtained a pole
mass of $M_c = 1.88 ^{+0.22}_{-0.13}$ GeV. This value resulted from
large corrections of the Coulomb potential to the ground state.

During the last years, several lattice analyses obtained the following
values for the ${\rm \MSb}$-mass:
\begin{eqnarray}
  \label{eq:o}
   m_{c}(m_{c}) &=& 1.59 \pm 0.28\ \mbox{GeV}\ \  \mbox{\cite{accflm:94}}\,,\nn\\
   m_{c}(m_{c}) &=& 1.33 \pm 0.08\ \mbox{GeV}\ \  \mbox{\cite{k:98}}\,, \nn\\
   m_{c}(m_{c}) &=& 1.73 \pm 0.26\ \mbox{GeV}\ \  \mbox{\cite{ggrt:99}}\,, \nn\\
   m_{c}(m_{c}) &=& 1.22 \pm 0.05\ \mbox{GeV}\ \ \mbox{\cite{bf:96,bf:97}}\,.
\end{eqnarray}
Whereas the results from \cite{k:98,bf:96,bf:97} are in good agreement
with this analysis, the investigations from \cite{accflm:94} and \cite{ggrt:99}
obtain higher masses. For the time-being, the results are not conclusive
and future lattice calculations for the charm mass might be 
of interest.
Since other methods reveal significant uncertainties in the determination
of the quark masses, QCD sum rules remain amongst the most precise
tools to extract these fundamental quantities.
                  
\bigskip \noindent
{\bf Acknowledgements}
The authors would like to thank Nora Brambilla, Antonio Pich and
Antonio Vairo for helpful and interesting discussions. Markus Eidem\"uller
thanks the Graduiertenkolleg ``Physikalische Systeme mit vielen
Freiheitsgraden'' at the University of Heidelberg for financial support.

\end{document}